\documentclass{PoS}
\usepackage{amsmath}

\title{NNLO corrections to semileptonic\\ and hadronic B decays}

\ShortTitle{NNLO corrections to semileptonic and hadronic B decays}

\author{\speaker{Guido Bell}%
         \thanks{I would like to thank Volker Pilipp for collaboration on part of the subjects presented in this
          article. This work was supported by the DFG Sonderforschungsbereich/Trans\-regio 9.}\\
        Institut f\"ur Theoretische Teilchenphysik,\\
        Universit\"at Karlsruhe,\\
        76128 Karlsruhe, Germany\\
        and\\
        Albert Einstein Center for Fundamental Physics,\\
        Institute for Theoretical Physics,\\
	University of Bern,\\
        3012 Bern,
        Switzerland\\
        E-mail: \email{bell@itp.unibe.ch}
}

\abstract{We review recent progress in the perturbative calculations for semileptonic and hadronic $B$ meson decays with an emphasis on the phenomenological implications of the next-to-next-to-leading order (NNLO) corrections. Specifically, we consider CP-averaged branching ratios of tree-dominated $B\to\pi\pi/\pi\rho/\rho\rho$ decays and the inclusive determination of $|V_{ub}|$ from $B\to X_u\ell\nu$.}

\FullConference{RADCOR 2009 - 9th International Symposium on Radiative
Corrections (Applications of Quantum Field Theory to Phenomenology) \\
                 October 25-30 2009\\
                 Ascona, Switzerland}

\begin{document}

\section{Introduction}

The $B$ physics programme continues to play a crucial role in testing the CKM mechanism of quark flavour mixing and in determining fundamental Standard Model parameters. In recent years accurate measurements of numerous $B$ physics observables have been realized, which calls for an equal improvement in refining the theoretical expectations within and beyond the Standard Model.

The main obstacle for precise theoretical predictions are the complicated strong-interaction effects encoded in the hadronic matrix elements. The QCD dynamics often simplifies considerably in the heavy quark limit $m_b \gg \Lambda_{QCD}$, which allows to establish factorization theorems that disentangle short- and long-distance effects. This separation provides the key for a systematic improvement of the theo\-retical predictions by computing higher order radiative corrections, which should be supplemented by similar progress in the determination of the remnant non-perturbative hadronic parameters.

Here we report on recent progress in the perturbative calculations for charmless hadronic and semileptonic $B$ meson decays. As a technical account of these calculations has already been presented in~\cite{Bell:2009rg}, we focus here on the phenomenological implications of the NNLO corrections.

\section{Hadronic B decays}

Most of the observables at current and future $B$ physics experiments are related to hadronic two-body decays. Among these time-dependent and direct CP-asymmetries in penguin-dominated decay modes are of particular phenomenological interest due to their sensitivity to New Physics.

The control of the strong-interaction dynamics in non-leptonic decays is obviously demanding. The factorization formula for the hadronic matrix elements of the operators in the weak effective Hamiltonian takes a twofold structure~\cite{QCDF},
\begin{eqnarray}
\langle M_1 M_2 | Q_i | \bar{B} \rangle
&\;\simeq\; &
F^{B M_1}(0) \; f_{M_2}
\int du \;T_{i}^I(u) \; \phi_{M_2}(u)
\nonumber \\
&&
+ \; \hat{f}_{B} \; f_{M_1} \; f_{M_2}
\int d\omega dv du \; T_{i}^{II}(\omega,v,u)
\; \phi_B(\omega) \; \phi_{M_1}(v) \;  \phi_{M_2}(u),
\end{eqnarray}
which consists of universal non-perturbative parameters (form factor $F^{B M}(q^2=0)$, decay constants $f_M$, light-cone distribution amplitudes $\phi_M$) and perturbative hard-scattering kernels $T_i^{I,II}$, that contain the short-distance dynamics of the flavour-changing quark transition. The latter are currently being worked out to NNLO, i.e.~at $\mathcal{O}(\alpha_s^2)$. While the full set of hard-scattering kernels from spectator scattering ($T_i^{II}$) for tree~\cite{NNLO:T2:tree} and penguin amplitudes~\cite{NNLO:T2:peng} is now available at NNLO, the ones related to the vertex corrections ($T_i^{I}$) are known to date for the tree amplitudes only~\cite{NNLO:T1:tree:GB,NNLO:T1:tree:AC}.

The NNLO calculation is particularly important for direct CP asymmetries that are first generated at $\mathcal{O}(\alpha_s)$. As in any perturbative calculation, it may thus help to reduce scale ambiguities of the leading contribution. The NNLO corrections may even change the pattern of CP asymmetries significantly, since they can potentially be enhanced by large Wilson coefficients (which is not possible at even higher orders since the NNLO terms already have the full complexity).

The current status of the NNLO calculation does not yet allow to discuss CP asymmetries. We may, however, already consider (CP-averaged) branching ratios of tree-dominated decay modes that do not depend significantly on the penguin amplitudes. As these observables are likely to be dominated by their Standard Model contributions, they may serve as important probes for our understanding of the strong-interaction dynamics in non-leptonic decays.

The NNLO analysis of the eleven tree-dominated $B\to\pi\pi/\pi\rho/\rho\rho$ decay modes has been presented in~\cite{Bell:2009fm} (for a similar analysis cf.~\cite{NNLO:T1:tree:AC}). In general colour-allowed decay modes turn out to be under much better theoretical control than colour-suppressed modes. Let us illustrate this point at the amplitude level: for the colour-allowed amplitude in the $\pi\pi$ channels one finds 
$\alpha_1(\pi\pi) = 1.013^{+0.023}_{-0.036} + (+0.027^{+0.025}_{-0.022})i$,
which is to be compared with
$\alpha_2(\pi\pi) = 0.195^{+0.134}_{-0.089} + (-0.101^{+0.061}_{-0.063})i$
for the colour-suppressed one. It is striking that the latter suffers from substantial theoretical uncertainties, which can be traced back to a strong (and unfortunate) cancellation between different terms in the perturbative expansion. This makes the real part of $\alpha_2$ particularly sensitive to the spectator scattering mechanism, which is normalized by the hadronic ratio $f_{\pi}\hat{f}_B/\lambda_BF_+^{B \pi}(0)$. The poor knowledge of the $B$ meson parameter $1/\lambda_B=\int_0^\infty d\omega/\omega\; \phi_B(\omega)$, in particular, makes the theoretical prediction of the colour-suppressed amplitude rather uncertain.

In order to test the QCD dynamics in hadronic decays it is useful to consider \emph{ratios} of decay rates rather than absolute branching fractions. Particularly suited are ratios that involve the differential semileptonic decay rate at maximum recoil,
\begin{align}
{\cal{R}}_{M_3}(M_1   M_2) =
\frac{\Gamma(\bar{B} \to M_1   M_2)}
{d\Gamma(\bar{B}^0\to M_3^+\ell^-\bar{\nu}_l)/dq^2|_{q^2=0}}.
\end{align}
Experimentally this requires to measure the semileptonic decay spectrum over a sufficiently large number of $q^2$-bins. Assuming specific parameterizations for the form factor shapes, the spectrum may then be extrapolated to $q^2=0$. At present this information is available for $B\to\pi\ell\nu$ decays~\cite{piellnu}, whereas the data on the $B\to\rho\ell\nu$ spectrum does not yet allow for an accurate extrapolation.

\begin{table*}[b!]
\begin{center}
\begin{tabular}{|r|cccc|} \hline
\hspace*{0cm}&
\hspace*{2cm}&\hspace*{2cm}&\hspace*{2cm}&\hspace*{2cm}
\\[-1.1em]
& ${\cal{R}}_\pi(\pi^-\pi^0)$
& ${\cal{R}}_\pi(\pi^+\pi^-)$
& ${\cal{R}}_\pi(\pi^0\rho^-)$
& ${\cal{R}}_\pi(\pi^+\rho^-)$
\\ [0.1em]
\hline
&&&&
\\[-1.2em]
Theory
& $0.70^{+0.12}_{-0.08}$
& $1.09^{+0.22}_{-0.20}$
& $1.71^{+0.27}_{-0.24}$
& $2.77^{+0.32}_{-0.31}$
\\[0.2em]
Experiment
& $0.81^{+0.14}_{-0.14}$
& $0.80^{+0.13}_{-0.13}$
& $1.57^{+0.32}_{-0.32}$
& $2.43^{+0.47}_{-0.47}$
\\[0.1em]
\hline
\end{tabular}
\end{center}
\vspace{-4.5mm}
\parbox{15.1cm}{\caption{\label{tab:slratios}
Ratios of hadronic and differential semileptonic decay rates in units of $\text{GeV}^2$.}}
\end{table*}

In Table~\ref{tab:slratios} we confront the NNLO prediction of the ${\cal{R}}_{\pi}$-ratios with experimental data. We stress that the theoretical predictions from Table~\ref{tab:slratios} are based on a default set of hadronic input parameters (specified in Table~I of~\cite{Bell:2009fm}), that is motivated by recent lattice and sum rule calculations. We see that the theoretical predictions are in good agreement with the data, which strongly supports the factorization assumption\footnote{The agreement is less pronounced for the ratio ${\cal{R}}_\pi(\pi^+\pi^-)$, which shows a much stronger dependence on the QCD penguin amplitude and hence on the specific input value for the weak phase $\gamma$. This ratio is thus not particularly suited to test the dynamics of the tree amplitudes.}. This is in particular true for the ratio ${\cal{R}}_\pi(\pi^-\pi^0)$, which does not depend on the QCD penguin amplitude and on weak annihilation contributions at all; it thus gives clean access to $|\alpha_1(\pi\pi)+\alpha_2(\pi\pi)|^2$~\cite{facttest}. Taking current data at face value, we may conclude that the colour-suppressed amplitude is somewhat enhanced, which may hint at a lower value of the $B$ meson parameter $\lambda_B\simeq~250$MeV (the default choice adopted in~\cite{Bell:2009fm} is $\lambda_B=(400\pm150)$MeV). It would be interesting to verify if this conclusion is supported by the according ratio in the $\rho$-sector. As long as the experimental information on the semileptonic $B\to\rho\ell\nu$ spectrum is absent, we may instead consider ratios of two hadronic decay rates,
\begin{align}
R(M_1 M_2/M_3 M_4) =
\frac{\Gamma(\bar{B} \to M_1 M_2)}
{\Gamma(\bar{B}' \to M_3 M_4)}.
\end{align}
The ratio $R(\rho_L^- \rho_L^0/\rho_L^+ \rho_L^-) \simeq |\alpha_1(\rho_L\rho_L)+\alpha_2(\rho_L\rho_L)|^2/2|\alpha_1(\rho_L\rho_L)|^2$ yields complementary information on the tree amplitudes from the $\rho$-sector\footnote{The subscript $L$ refers to the longitudinal polarization.}. One should keep in mind, however, that this ratio receives corrections from the QCD penguin amplitude and from weak annihilation in contrast to the semileptonic ratio ${\cal{R}}_{\rho}(\rho_L^- \rho_L^0)$. From the numbers in Table~\ref{tab:hadratios} we infer that the NNLO prediction is again found to be smaller than the experimental value, which supports the hypothesis of enhanced colour-suppressed amplitudes and hence a lower value of $\lambda_B$.

\begin{table*}[b!]
\begin{center}
\begin{tabular}{|r|c|cc|} \hline
\hspace*{0cm}&
\hspace*{2.5cm}&\hspace*{2.5cm}&\hspace*{2.5cm}
\\[-1.1em]
& $R(\rho_L^- \rho_L^0/\rho_L^+ \rho_L^-)$
& $R(\pi^0 \rho^0/\pi^0 \pi^0)$
& $R(\pi^0 \rho^0/\rho_L^0 \rho_L^0)$
\\ [0.1em]
\hline
&&&
\\[-1.2em]
Theory
& $0.65^{+0.16}_{-0.11}$
& $1.50^{+1.70}_{-1.32}$
& $1.17^{+0.45}_{-0.43}$
\\[0.2em]
Experiment
& $0.89^{+0.14}_{-0.14}$
& $1.29^{+0.36}_{-0.36}$
& $2.90^{+1.45}_{-1.45}$
\\[0.1em]
\hline
\end{tabular}
\end{center}
\vspace{-4mm}
\parbox{15.1cm}{\caption{\label{tab:hadratios}
Ratios of two hadronic decay rates.}}
\end{table*}

Let us finally comment on the colour-suppressed modes, which are more complicated due to their strong dependence on hadronic input parameters. In contrast to the colour-allowed modes, it is in particular \emph{not} possible to reduce these uncertainties by considering semi\-leptonic ${\cal{R}}_{M}$-ratios since their dependence on $|V_{ub}|^2 |F_+^{BM}(0)|^2$ is weak. One may instead try to resolve the correlation among the theoretical uncertainties by considering hadronic ratios of two colour-suppressed modes. As can be seen in~Table~\ref{tab:hadratios}, this does unfortunately not lead to an improvement for the ratio that involves the $\pi^0\pi^0$ decay mode\footnote{Let us emphasize that the agreement between the central values is accidental for this ratio, since the theoretical prediction for the individual branching ratios are quite below the experimental data (the numbers can be found in~\cite{Bell:2009fm}).}. This is different for the ratio $R(\pi^0 \rho^0/\rho_L^0 \rho_L^0)$, which is less contaminated by the QCD penguin amplitudes. Consequently, the dependence on $\lambda_B$ and, somewhat accidentally, the one from the modelled power corrections drop out to a large extent. A more precise experimental value for this ratio may therefore give further insight into the role of power corrections in non-leptonic decays. The dynamics of the colour-suppressed amplitudes, however, should be probed with the cleaner ratios ${\cal{R}}_{\pi}(\pi^- \pi^0)$, ${\cal{R}}_{\rho}(\rho_L^- \rho_L^0)$, ${\cal{R}}_{\rho}(\pi^- \rho^0)$ and ${\cal{R}}_{\pi}(\pi^0\rho^- )$.

\section{Semileptonic B decays}

Semileptonic $b\to u$ decays provide a measure of the CKM matrix element $|V_{ub}|$. The current discrepancy between inclusive and exclusive determinations calls for further progress on both sides. Here we report on the NNLO calculation for inclusive $B\to X_u\ell\nu$ decays in the so-called BLNP approach~\cite{BLNP}.

The theoretical description of inclusive $B\to X_u\ell\nu$ decays is complicated by the fact that experimental measurements have to introduce kinematical cuts to suppress the $B\to X_c\ell\nu$ background. This restricts the experimental information to the shape-function region in which the hadronic final state has large energy $E_X \sim m_b$ but moderate invariant mass $p_X^2 \sim m_b \Lambda_{QCD}$. In this region of phase space a factorization formula for the structure functions has been put forward \cite{Korchemsky:1994jb}
\begin{eqnarray}
W_i &\;\simeq\;&
 H_{i} \,
\int d\omega \; J(p_\omega^2) \; S(\omega),
\end{eqnarray}
which contains an universal non-perturbative quantity, the shape function $S$, and two perturbatively calculable objects, hard coefficient functions $H_{i}$ and a jet function $J$, that encode the short-distance effects. The NNLO calculation of the latter is now complete. While the two-loop corrections to the jet function have been worked out in~\cite{Becher:2006qw}, the hard coefficient functions have recently been computed to NNLO by various groups~\cite{BtoXuellnu}. The latter calculation required to match the flavour-changing $V-A$ current from QCD onto soft-collinear effective theory~\cite{SCET}. It has further been generalized to the tensor current, which finds applications in electroweak penguin decays~\cite{BBHL}.

The numerical impact of the NNLO corrections on the inclusive determination of $|V_{ub}|$ has recently been analyzed in~\cite{Greub:2009sv}. Starting from the two-loop expressions for the hard and jet functions, the authors implemented the renormalization group improvement and a specific model for the shape function. From their analysis of partial decay rates, they conclude that the NNLO corrections can be significant. This statement, however, depends on the choice of the (arbitrary) matching scale $\mu_i\sim(m_b \Lambda_{QCD})^{1/2}$. For $\mu_i= 1.5$~GeV, which was the default choice in the earlier BLNP analysis~\cite{BLNP}, the NNLO corrections are found to be important; they typically lower partial decay rates by about $15-20\%$ while at the same time reducing the perturbative uncertainties.

\begin{table*}[b!]
\begin{center}
\begin{tabular}{|lccc|} \hline
\hspace*{0cm}&
\hspace*{2.5cm}&\hspace*{2.5cm}&\hspace*{2.5cm}
\\[-1.1em]
Exp. & Method
& $|V_{ub}|~[10^{-3}]$
& $|V_{ub}|~[10^{-3}]$
\\[-0.1em]
& 
& NLO
& NNLO
\\ [0.1em]
\hline
&&&
\\[-1.2em]
BABAR
& $E_l>2.0$~GeV
& $3.97\pm0.22^{+0.37+0.26}_{-0.23-0.25}$
& $4.30\pm0.24^{+0.26+0.28}_{-0.20-0.27}$
\\[0.2em]
BELLE
& $M_X<1.7$~GeV
& $3.55\pm0.24^{+0.22+0.21}_{-0.13-0.19}$
& $3.87\pm0.26^{+0.21+0.21}_{-0.13-0.19}$
\\[0.2em]
BABAR
& $P_+<0.66$~GeV
& $3.30\pm0.23^{+0.27+0.25}_{-0.16-0.22}$
& $3.55\pm0.24^{+0.19+0.24}_{-0.13-0.21}$
\\[0.1em]
\hline
\end{tabular}
\parbox{12.1cm}{\caption{\label{tab:Vub}
Values of $|V_{ub}|$ deduced from different experimental measurements of partial $B\to X_u\ell\nu$ decay rates (the numbers are taken from~\cite{Greub:2009sv}).}}
\end{center}
\end{table*}

In their determination of $|V_{ub}|$ the authors combine the NNLO prediction of the leading term in the heavy quark expansion with known power corrections up to $\mathcal{O}(1/m_b^2)$. From the experimental information on partial decay rates, which are based on different types of experimental cuts (lepton energy $E_l$, hadronic invariant mass $M_X$, hadronic variable $P_+ = E_X - |\vec{P}_X|$), they deduce sample values for $|V_{ub}|$. Some of their results are collected in Table~\ref{tab:Vub}, which illustrate that the central values are shifted significantly at NNLO. Concerning the error estimate, the first one reflects the experimental uncertainty, while the improvement on the perturbative uncertainty can be seen in the second one. One further infers from the last error that the numerical value of the $b$-quark mass, which enters certain moment constraints of the shape-function model, has a large impact on the determination of $|V_{ub}|$. Given that the NNLO calculation has increased the discrepancy with the exclusive determination, further theoretical progress on the treatment of shape function effects is highly desirable (cf., for instance,~\cite{Ligeti:2008ac} for an alternative implementation that is not based on a specific model).

\end{document}